
\def\gam{\gamma}
\def\hp{H^+}
\def\hm{H^-}
\def\hpm{H^\pm}
\def\mhpm{m_{\hpm}}
\def\anti{\overline}
\def\mz{m_Z}
\def\gev{~{\rm GeV}}
\def\tev{~{\rm TeV}}
\def\pbi{~{\rm pb}^{-1}}
\def\fbi{~{\rm fb}^{-1}}


\def\hl{h^0}
\def\ha{A^0}
\def\hh{H^0}
\def\mhl{m_h}
\def\mhh{m_H}
\def\mha{m_A}
\def\mt{m_t}
\def\rta{\rightarrow}
\def\tanb{\tan\beta}
\def\lplm{l^+l^-}

\input phyzzx
\Pubnum={$\caps UCD-92-27$\cr}
\date{November, 1992}
\def\mstop{m_{\wtilde t}}

\titlepage
\baselineskip 0pt
\hsize=6.5in
\vsize=8.5in
\centerline{{\bf SEARCHING FOR}}
\vskip .05in
\centerline{{\bf THE MINIMAL SUPERSYMMETRIC MODEL HIGGS BOSONS}
\foot
{To appear in {\it Proceedings of the 1992 Division of Particles
and Fields Meeting}, Fermilab, Nov. 11-15 (1992), World Scientific.}
}
\vskip .075in
\centerline{John F. Gunion}
\vskip .075in
\centerline{\it Davis Institute for High Energy Physics}
\centerline{\it Department of Physics, U.C. Davis, Davis CA 95616}
\vskip .075in
\centerline{ABSTRACT}
\vskip .075in
\centerline{\Tenpoint\baselineskip=12pt
\vbox{\hsize=12.4cm
\noindent A brief overview of the prospects for detecting the
Higgs bosons of the Minimal Supersymmetric Model
at future colliders is presented.}}

Probing the Higgs sector is
the fundamental mission of future high energy colliders.
The Minimal Supersymmetric Model (MSSM) is the simplest supersymmetric
extension of the Standard Model (SM), and contains five physical
Higgs bosons:  the light $\hl$, and heavier $\hh$, $\ha$, and $\hpm$.
We review our ability to detect these Higgs bosons at LEP-II, SSC/LHC and
a next linear $\epem$ collider (NLC) with $\sqrt s=500\gev$. In the latter
case,
we examine both direct $\epem$ collisions and Higgs production via
collisions of back-scattered laser beams.
\REF\detmssm{J.F. Gunion, preprint UCD-92-20 (1992), to appear
in {\it Perspectives in Higgs Physics}, ed. G. Kane, World Scientific.}
See Ref.~[\detmssm] for details, plots and references.

As is well-known, the MSSM Higgs sector at tree-level is
completely determined by the choice of just two parameters, conventionally
taken as $\mha$ and $\tanb=v_2/v_1$.
To good approximation, the only additional parameters needed to determine the
one-loop radiative corrections to the Higgs bosons' masses and couplings
are $\mt$ and the stop mass, $\mstop\,$; in particular,
$\left(\mhl^{max}\right)^2\sim \mz^2+const.\times \mt^4\ln(\mstop^2/\mt^2)$
for large $\mha$ and $\tanb$.
(However, to determine the decay patterns of the Higgs bosons
the masses of neutralinos and charginos are also required.)
The dependence of Higgs masses, couplings and decays on the MSSM parameters
are sufficiently constrained that we can outline the regions
of parameter space for which detection of a given Higgs boson is
possible at any given collider.  A summary of the most important
results obtained after investigating signals and backgrounds is given
in the following sections.

\smallskip
\noindent{\bf 1. LEP-II}
\smallskip

Certainly the most important goal of LEP-II will be to detect the $\hl$
if it exists.  In the most likely scenario where
$\mha\gsim 2\mz$, $\hl$ would be searched for in the $\epem\rta Z^*\rta Z\hl$
mode. Since the LEP-II energy is likely to be in the vicinity of $2\mz$,
and since $\mhl$ can be larger than $\mz$, our ability to find
the $\hl$ at LEP-II will be extremely sensitive to $\sqrt s$, $\tanb$,
$\mt$ and $\mstop$.  At large $\mha$,  consider fixed $(\sqrt s,\tanb)$
and plot $Z\hl$ discovery contours in $\mt$--$\mstop$ parameter space.
(100 events for $L=500\pbi$ are adequate.) As expected,
$\hl$ detection becomes impossible if either $\mt$ or $\mstop$
is too large. The larger $\tanb$ and/or the smaller $\sqrt s$,
the lower the values of $\mt,\mstop$ beyond which discovery is impossible.
The most rapid variation in contour location occurs when $\tanb \gsim 15$ and
$\sqrt s$ is increased from 190 GeV to 200 GeV. If $\mt=135\gev$
and $\tanb=20$, detection of the $\hl$ is possible for all $\mstop\lsim 1\tev$
at $\sqrt s=200\gev$, but becomes {\it impossible} for $\mstop\gsim 200\gev$
if $\sqrt s=190\gev$.  A LEP-II energy as low as $175\gev$ would
make $\hl$ detection essentially impossible for any values
of $\mt$ and $\mstop$ at $\tanb=20$. In the opposite extreme,
$\sqrt s=240\gev$ would guarantee
$\hl$ discovery for any $\mt\lsim 200\gev$ and
$\mstop\lsim 1\tev$ even if $\tanb \sim 20$.
Once $\mt$ is known, the precise extent to which the LEP $\sqrt s$
should be pushed to guarantee $\hl$ discovery for reasonable
values of $\tanb\lsim 20$ and $\mstop\lsim 1\tev$ will be known.

\smallskip
\noindent{\bf 2. NLC-500}
\smallskip

Should the $\hl$ not be discovered at LEP-II, an NLC with $\sqrt s=500\gev$
will certainly be able to find it, provided an integrated luminosity
of $L=10\fbi$ is accumulated. In the most likely scenario where
$\mha\gsim 100\gev$, the $\epem\rta Z^*\rta Z\hl$ and $WW$-fusion
$\epem\rta\nu\anti\nu W^*W^*\rta\nu\anti\nu\hl$
processes will both yield at least
100 events for all $\tanb$ values, regardless of the values of $\mt$
and $\mstop$.  However, detection of the $\hh$, $\ha$, and $\hpm$ is
not guaranteed.  In fact, for $\mha\gsim 120\gev$ the coupling constant
relations of the MSSM imply that the only production processes
for these Higgs bosons with potentially large
rates are the $\epem\rta Z^*\rta \hh\ha$
and $\epem\rta Z^*\rta \hp\hm$ pair-production reactions.
When not phase-space suppressed, these reactions yield at least 100 events
for all $\tanb$ once $\mha\gsim 100\gev$.  However, if we recall
that $\mhh\sim\mha\sim\mhpm$ for $\mha\gsim 2\mz$, it is clear that
both reactions become kinematically disallowed for $\mha+\mhh\sim 2\mhpm
\sim \sqrt s$.  At $\sqrt s=500\gev$, the kinematical suppression
is such that the largest $\mha\sim\mhh\sim\mhpm$ value that can
be probed at the NLC will be between $200$ and $210\gev$, \ie\
roughly $0.4\sqrt s$.

\smallskip
\noindent{\bf 3. SSC/LHC}
\smallskip

Of course, even before an NLC will be available the SSC and LHC will
both be running.  What will be their ability to detect the MSSM Higgs
bosons?  The answer depends critically on many parameters.  In order
to restrict the discussion, I adopt a fixed top quark mass of $\mt=150\gev$,
an integrated SSC luminosity of $L=30\fbi$ (results for the LHC
with $L=100\fbi$ are very similar), and consider three scenarios for
other relevant parameters: a) $\mstop=1\tev$ and all ino masses $\gsim
200\gev$; b) $\mstop=1\tev$ and ino masses
specified by $M=200\gev$ and $\mu=100\gev$; c) $\mstop=300\gev$
and ino masses as in b).
(The large-$\tanb$ ino masses for $M=200\gev$
and $\mu=100\gev$ are: $\wtilde \chi_i^0$ --- 59, 114, 117, 237 GeV;
$\wtilde \chi_i^+$ --- 82, 237 GeV. Even at low $\tanb$, $m_{\wtilde\chi_1^0}$
is above the cosmologically motivated bound quoted by the PDG.)
In scenarios a) and b) the radiative
corrections to the MSSM Higgs masses (especially $\mhl$ and the lower
values of $\mhh$) are large.  In scenario c), these radiative
corrections are much smaller. In scenario a) Higgs decays to superpartner
particles do not occur for the parameter region to be discussed,
whereas in scenarios b) and c) Higgs decays to ino pairs are of considerable
importance (sometimes dominant) for the $\hh$, $\ha$ and $\hpm$
in the higher mass region.

We discuss detection modes involving SM final
state particles for which backgrounds have been thoroughly studied
and specific detection criteria are known at a high level of confidence.
First, there is $W^*\rta WH$ plus $gg\rta t\anti t H\rta WH X$
($H=\hl,\hh,\ha$)
followed by $H \rta\gam\gam$ and $W\rta l\nu$ --- the `$l\gam\gam$' mode.
Second, there is $gg\rta H\rta ZZ~{\rm or}~ZZ^*\rta \lplm\lplm$ ---
the `$4l$' mode. Finally, there is detection of $t\rta \hpm b$ decays.
Let us examine the feasibility of the various modes in terms of the
$\mha$--$\tanb$ parameter space, other masses and parameters being
determined by the MSSM relations. We consider only $\mha\leq 400\gev$
and $0.5\leq\tanb\leq 20$. Recent limits on $BR(b\rta s \gam)$
imply a lower limit on $\mhpm$ such that
$\mha$ values below about $100\gev$ are probably ruled out.
This implies $t\rta\hpm b$ will not be possible at $\mt=150\gev$.
For $\mha\gsim 100\gev$, $W\hh$ detection in the $l\gam\gam$ mode
is also not possible.
Further, $\mhl$ is too light at $\mt=150\gev$ for $\hl\rta 4l$
detection, and $\hh\rta 4l$ detection is possible only in a very limited region
--- $50\lsim \mha\lsim 2\mt$ when $\tanb\lsim 3.5-7$ in scenario a),
decreasing substantially in b), and essentially gone in c).
$\ha$ detection in any mode is impossible unless $\tanb\lsim 1$.
However, more or less independent
of scenario, the $W\hl\rta l\gam\gam X$ mode at the SSC {\it will} be viable
for (roughly) $\mha\gsim 160\gev$ at all $\tanb$.

In all three scenarios the SSC will not be able to
detect any MSSM Higgs boson (in the modes discussed)
between $\mha\sim 100$ and $\mha\sim 160\gev$ (narrower
at small $\tanb$). Over some portion of this region,
LEP-II (at $\sqrt s=200\gev$) will be able to detect $\hl Z$ events.
For scenarios a) and b) this region is limited to $\tanb\lsim 12-9$
for $\mha\sim 100 - 150\gev$ --- the large
$\mstop$ implies $\mhl$ is too large at higher $\tanb$ values.
However, in scenario c), $\mhl$ is always sufficiently light
that $\hl Z$ detection
will be possible at LEP-II for $\mha\gsim100\gev$ for all $\tanb$.

\smallskip
\noindent{\bf 4. Back-Scattered Laser Beams at the NLC}
\smallskip

Since only
the $\hl$ will be easily discovered at LEP-II and/or SSC/LHC, it becomes
crucial
to see if the reach of NLC-500 in $\ha$, $\hh$ or $\hpm$ mass can be
extended beyond $\sim 0.4\sqrt s$. We find
that $\gam\gam$ collisions of back-scattered laser
beams can push the discovery region for the $\ha$ and $\hh$ to $\sim 0.8
\sqrt s$, assuming an `effective luminosity' of $20\fbi$.
For instance, the viable channels in the case of the $\ha$ are:
$\gam\gam\rta\ha\rta Z\hl$ or $ b\anti b$, for $\mha\lsim 2\mt$;
and $\gam\gam\rta\ha\rta t\anti t$, for $\mha\gsim 2\mt$.  The
$\gam\gam\rta b\anti b$ and $t\anti t$ backgrounds can be suppressed by
appropriate cuts and adequate polarization for the colliding $\gam$'s.
Extension to $\ha$ masses of order $0.8\sqrt s$ is only possible
for small to moderate values of $\tanb$.  By $\tanb=20$, the only useful
channel is $\gam\gam\rta\ha \rta b\anti b$; event
rates are adequate for $\mha\lsim 200 - 250\gev$
(at $L_{eff}=20\fbi$).

\smallskip
\noindent{\bf 5. Conclusions}
\smallskip

Combining LEP-II and SSC/LHC,
detection of the $\hl$ is almost guaranteed, but detection of the $\ha$,
$\hh$, and $\hpm$ is likely to require the use of other (less background
free) final state channels.  The utility of these other modes
remains uncertain at this point in time.
The ability to detect the $\ha$ and $\hh$ is much greater at the NLC.

\smallskip\noindent{\bf 5. Acknowledgements}
\smallskip
I would like to thank my collaborators for their many
contributions, especially L. Orr, H. Haber, and M. Barnett.
This work has been supported in part by the Department of Energy.
\smallskip
\noindent{\bf 6. References}
\smallskip
\point J.F. Gunion, preprint UCD-92-20 (1992), to appear
in {\it Perspectives in Higgs Physics}, ed. G. Kane, World Scientific.
\end